\documentclass[aps,twocolumn,showpacs,amsmath,floatfix,superscriptaddress]{revtex4}

\usepackage{graphicx}
\usepackage{dcolumn}
\usepackage{bm}
\usepackage{accents}
\usepackage{xcolor}
\begin{document}

\title{Topological Protected Dirac Cones in Compressed Bulk Black Phosphorus}

\author{Ruixiang Fei, Vy Tran and Li Yang}

\affiliation{Department of Physics, Washington University in St.
Louis, St. Louis, MO 63130, USA}

\date{}

\begin{abstract}
Using the $k\cdot p$ theory and first-principles simulations, we
report that applying a moderate pressure ($>$ 0.6 GPa) on black
phosphorus can diminish its band gap and produce one-dimensional
and even two-dimensional (2D) Dirac cones, distinguishing this
material for use in novel non-compound topological insulators.
Similar to topological insulators, these 2D Dirac cones result
from two competing mechanisms: the unique linear band dispersion
tends to open a gap via a ``pseudo spin-orbit'' coupling, while
the band symmetry requirements preserve the material's gapless
spectrum. Moreover, these unique Dirac cones are bulk states that
do not require time-reversal symmetry, thus they are robust even
in the presence of surface or magnetic perturbations. Ultimately,
we show that our predictions can be detected by the material's
unusual Landau levels.
\end{abstract}

\maketitle

To date Dirac materials, such as graphene \cite{2005Geim,
2005Zhang}, topological insulators (TIs) \cite{2009Hasan,
2009Shen, 2007Mele} or spin Hall insulators (SHIs)
\cite{2007SCZhang, 2004SCZhang}, topological crystalline
insulators (TCIs) \cite{2011Fu, 2012Tanaka} and Weyl semimetals
(WSs) \cite{2012Young, 2014Chen}, have drawn tremendous research
interest. Graphene hosts two-dimensional (2D) Dirac fermions
\cite{2009Neto,2011Sarma}; TIs are materials with a bulk energy
gap but possess gapless 2D (1D) Dirac surface states that are
protected by time-reversal symmetry \cite{2011Qi, 2010Kane}; TCIs
exhibit metallic surface states protected by the mirror symmetry
of the crystal \cite{2012Tanaka}; WSs, which are protected by
crystal symmetries, exhibit energy overlaps that occur only at a
set of isolated points in momentum space where the linearly
dispersed bands cross at the Fermi level \cite{2012Young}. The
impact and implications of these materials are difficult to
overstate. The discovery of the anomalous Hall effect
\cite{2011Niu, 2013Xue}, the fractional Hall effect \cite{2009Du},
chiral anomaly \cite{2012Zyuzin, 2014Burkov}, and Majorana
fermions \cite{2008Fu, 2009Beenakker}, to name a few, have
inspired enormous efforts in search for new Dirac materials.

Recently a new type of 2D semiconductor, few-layer black
phosphorus (BP) (phosphorene), has been fabricated
\cite{2014Zhang, 2014ye, 2014Xia}. It exhibits promising carrier
mobilities and unique anisotropic transport, optical, and thermal
properties \cite{2014Zhang, 2014ye, 2014Xia, 2014Fei, 2014Yang,
2014Yangnano, 2014Wang}. In particular, few-layer BP posses a
unique band structure, whose dispersion is nearly linear along the
armchair direction but parabolic along other directions
\cite{2014vy, 2014Neto}. An obvious idea follows from this unique
structure: if one can significantly reduce the band gap or even
achieve the band inversion, novel features may result, e.g., the
formation of Dirac cones. However, the band gap of monolayer and
few-layer BP is significant (up to 2.2 eV) \cite{2014Yang,
2014sts, 2014Wang}, making it impractical to close. Alternatively,
bulk black phosphorus has a much smaller band gap, only around 300
meV \cite{2014Yang, bulkbp}, which is more tractable.

In this work, we show that the $k\cdot p$ Hamiltonian of bulk BP
can be written as the free-electron-gas Hamiltonian with
off-diagonal terms for describing interband interactions, which
are of the same form as the Rasshba and Dressehaus spin-orbit
coupling (SOC). Thus we call this off-diagonal interaction the
"pseudo spin-orbit" coupling (PSOC). If the band gap is diminished
and the band inversion is realized, PSOC tends to open a gap in
the energy spectrum. However, similar to TIs, the band symmetry
requires band connections. Competition between PSOC and band
symmetry causes the Fermi surface to undergo a topological
transition from a sphere to a ring of 2D Dirac cones. Using
first-principles simulations, we show that pressure can be an
efficient tool to produce these 2D Dirac cones and even 1D Dirac
cones, while also tuning the characteristic Fermi velocity.
Interestingly, because our predicted Dirac cone feature in
compressed BP does not require the time reversal symmetry, it is
not destroyed by magnetic fields. Thus we provide a means for
detecting this unique electronic structure by demonstrating the
unusual energy scaling law exhibited by Landau levels.

\begin{figure}
\centering
\includegraphics[scale=0.25]{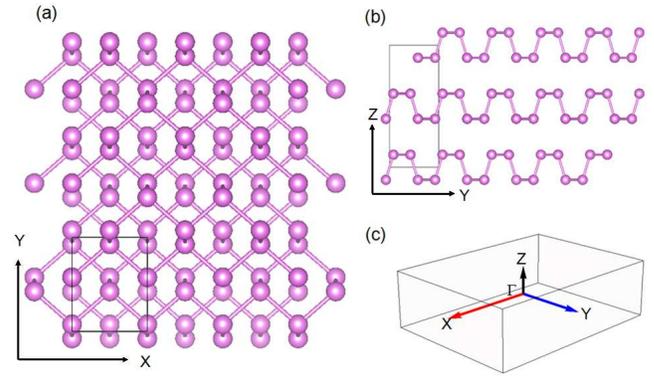}
\caption{\label{FIG. 1} (Color online) (a) Top view of the atomic
structure of bulk BP. The unit cell and lattice vectors are
marked. The zigzag and armchair directions are illustrated as
well. (b) Side view of BP. (c)The first BZ.}\label{Fig_1}
\end{figure}

The atomic structure of bulk BP is plotted in Figures 1 (a) and
(b). Here we use a simple orthorhombic conventional unit cell,
which is twice the size of the primitive cell. The corresponding
first Brillouin Zone (BZ) is shown in Figure 1 (c). The
double-cell procedure simply folds the first BZ and does not
affect any essential physics; this is convenient for presenting
the topology of the Dirac cones. The bulk BP atomic structures is
fully relaxed according to the forces and stresses calculated
using density functional theory (DFT) with the Perdew, Burke and
Ernzerhof (PBE) functional associated with van der Waals
corrections \cite{1996PBE, 2004Grimme}. Because DFT is known to
underestimate the band gap, we use a hybrid functional theory
(HFT), HSE06 \cite{2003HSE, 2006HSE}, to calculate electronic band
structures, which is in good agreement with the GW calculations
and experimental measurements \cite{2014Yang, bulkbp} of bulk BP.

\begin{figure}
\includegraphics[scale=0.45]{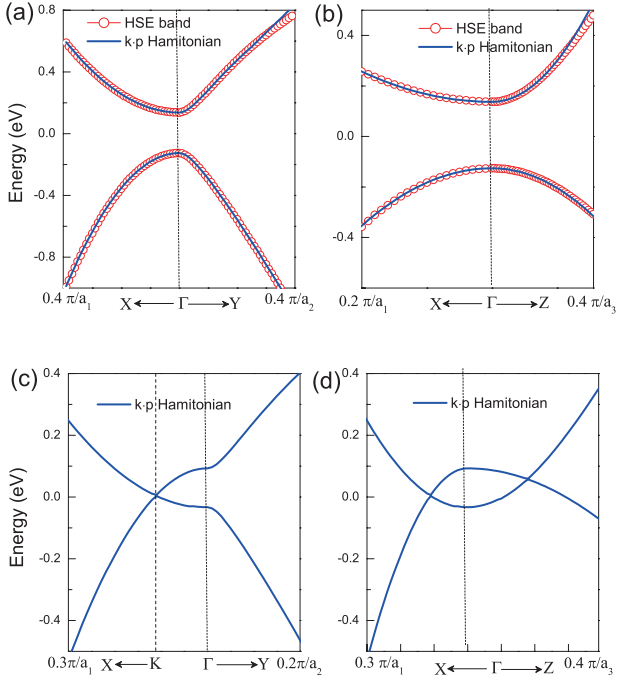}
\caption{(Color online)(a) and (b) are the band structures of
intrinsic BP. The red dots are from the HSE06 simulation and the
blue line is from the $k\cdot p$ calculations. (c) and (d) are the
band structures calculated from the $k\cdot p$ calculations under
the band-inversion condition. The Fermi level is set to be zero.}
\label{Fig_2}
\end{figure}

We begin with the widely used $k\cdot p$ model to capture the
essential BP's electronic structure. The $D_{2h}$ point group
invariance allows us to describe BP using a two-band model similar
to the one used for monolayer BP \cite{2014Chang, 2014Ezawa}. Thus
the low-energy effective Hamiltonian is given by
\begin{equation}
\begin{split}
&H= \left(
\begin{array}{cc}
E_c+\alpha k_x^2 +\beta k_y^2 +\gamma k_z^2 & v_f k_y \\
v_f k_y & E_v-\lambda k_x^2 -\mu k_y^2 -\nu k_z^2,
\end{array}
\right)
\end{split}
\end{equation}
where $k_x$, $k_y$, and $k_z$ are components of the crystal
momentum along the $x$, $y$, and $z$ directions, respectively. The
conduction band and the valance band energies, $E_c=0.15 \ eV$ and
$E_v=-0.15 \ eV$, are set to fit the bulk band gap (300 meV).
$\alpha$, $\beta$, $\gamma$, $\lambda$, $\mu$, and $\nu$ are
fitted parameters that describe the band edge curvature, and $v_f$
is the Fermi velocity of the nearly linear band dispersion along
the armchair ($y$) direction. By appropriately choosing these
parameters, as shown in Figures 2 (a) and (b), the $k\cdot p$
result nearly perfectly matches that of the first-principles HSE06
calculations of intrinsic BP.

The special characteristics of Eq. 1 are the off-diagonal terms of
the effective Hamiltonian, which produce the unique linear
dispersion along the armchair ($y$) direction, as shown in Figure
2 (a). In fact, This off-diagonal interaction can be split into
two terms
\begin{equation}
\begin{split}
&H_{off}=H_{DSOC}+H_{RSOC}\\
&=1/2\ v_f (k_x\sigma_y+k_y\sigma_x)-1/2\ v_f
(k_x\sigma_y-k_y\sigma_x),
\end{split}
\end{equation}
where $\sigma_x$ and $\sigma_y$ are Pauli matrices. Interestingly,
these two terms ($H_{DSOC}$ and $H_{RSOC}$) have the forms of the
Dressehaus and Rashba SOC \cite{2004Sarma}, respectively.
Realistically, Dressehaus and Rashba SOC cannot exist in intrinsic
bulk BP, because its lattice structure possesses the inversion
symmetry. Hence we call these interactions PSOC. Because of their
similar formalisms, we expect equivalent "SOC effects" to appear
in bulk BP. In particular, considering the dramatic effect of SOC
in TIs, it is natural to expect that diminishing or inverting the
band gap in bulk BP will produce novel electronic characteristics.

\begin{figure}
\centering
\includegraphics[scale=0.3]{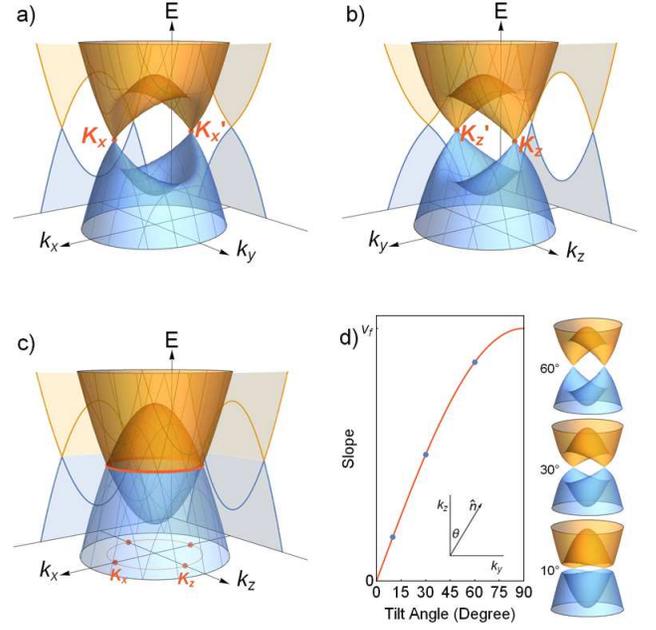}
\caption{(Color online) The 3D band structures of BP of the X-Y
plane (a), the Y-Z plane (b), and the X-Z plane (c). The band
overlap in (c) forms a ring structure. (d) The evolution of the
band structure from the X-Z plane to the X-Y plane via changing of
the tilted angle.} \label{Fig_3}
\end{figure}

It is easy to mimic the band inversion in the $k\cdot p$ model;
one simply assigns the valence band edge energy $E_v$ a higher
value than the conduction band edge energy $E_c$ in Eq. 1.
Surprisingly, instead of producing simple band overlaps, we
observe an unusual topological transition at the Fermi surface. In
Figures 2 (c) and (d), the band crossing at the $\Gamma$ point is
opened by the PSOC, while the band crossing is preserved at the K
point in the $\Gamma-X$ direction. As a result, 2D graphene-like
Dirac cones are formed and the Schematic 3D plots of the
electronic structure are presented in Figures 3 (a), (b), and (c)
for three specific cutting planes, X-Y, Y-Z and X-Z, respectively.
The linear band dispersions and Dirac cones are observed in both
X-Y and Y-Z planes and a ring-like band crossing occurs in the X-Z
plane. Actually, we can regard the Dirac cones observed in Figure
3 (a) and (b) are those points marked in the ring structures of
Figure 3 (c). In this sense, any point on the ring of Figure 3 (c)
corresponds to a Dirac cone in other 2D planes. In Figure 3 (d),
we show that this ring-like band crossing continuously evolves to
a pair of Dirac cones by varying the tilting angles from the X-Z
plane to the Y-Z plane. In other words, the set of tilting angles
that do not exhibit these Dirac cones is of measure zero, and thus
our findings are robust.

\begin{figure}
\centering
\includegraphics[scale=0.6]{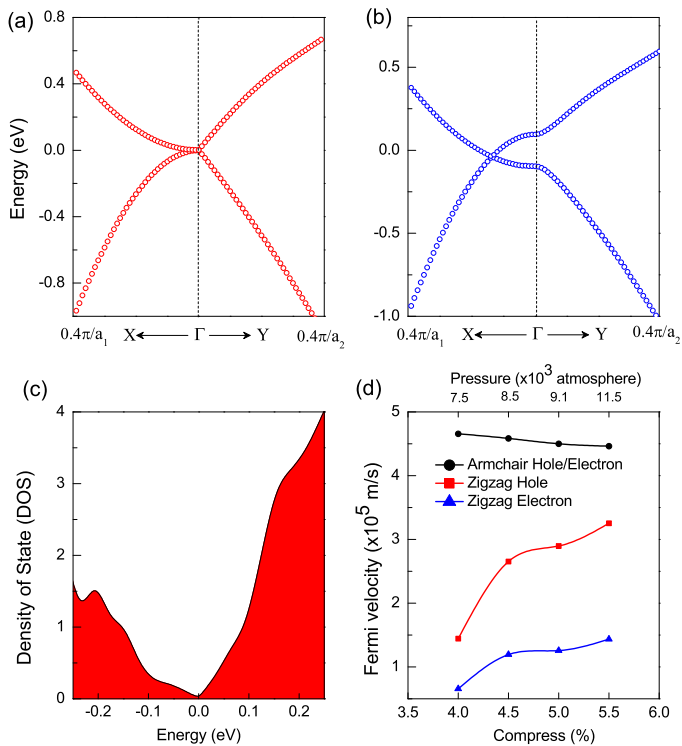}
\caption{(Color online)(a) The HSE06-calculated band structure of
BP under the critical pressure (0.6 GPa). (b) The HSE06-calculated
band structure of BP under the 0.9 GPa pressure. (c) The
Anisotropic fermi velocity of electrons and holes of 2D Dirac
cones varied with the pressure.} \label{Fig_4}
\end{figure}

The mechanisms responsible for forming the ring of Dirac cones are
similar to those in TIs, but there are some crucial differences.
First, similar to the band inversion and gap-opening that occurs
in TIs due to SOC interactions, the gap-opening at the $\Gamma$
point of Figure 2 (c) from the PSOC of states residing on the
linear band along the $\Gamma-Y$ direction. Second, the gap of BP
is not completely opened; the valence and conduction bands still
touch at the $K$ point (Figure 2 (c)) or by the ring (Figure 3
(c)). This gapless feature is guaranteed by a symmetry constraint
based on two facts: the bottom conduction band is formed by what
was originally top valence band prior to the band inversion; the
parity symmetries of the top of the valence band and the bottom of
the conduction band are inverse to each other, which is evidenced
by the bright dipole optical transitions \cite{2014Yang}.
Therefore, as shown in Figures 2 (c) and 3 (c), contact between
the conduction and valence band is required in order to provide a
continuous symmetry evolution. This mechanism is similar to the
one responsible for forming gapless surface states in TIs
\cite{2009Zhangh}. However, there are no surface states in our
calculated bulk BP, thus it is the bulk valence and conduction
bands that must touch. As a result, these two factors force the
formation Dirac cones. Moreover, as proposed by a recent work
\cite{2014Liu}, further breaking the inversion symmetry and
including the spin-orbit coupling in few-layer BP will open the
bulk band gap and produce surface topological states. Finally,
another special characteristic of these bulk Dirac cone states in
BP is that they are robust even in the face of broken time
reversal symmetry. Therefore, these Dirac cones can uniquely be
observed in experiments that employ magnetic fields and are
potentially useful for magnetic devices.

Following the above analysis of the $k\cdot p$ model and beyond 2D
Dirac cones, 3D Weyl points could be achieved if an extra
direction, such as the $x$ or $z$, also exhibits the off-diagonal
linear interband interaction (PSOC) term. Therefore, one may
search for WSs by looking for realizing the band inversion of the
materials that exhibit nearly linear dispersions along two
orthogonal directions.

Beyond the $k\cdot p$ model, an obvious question is how one can
practically reduce the band gap or achieve band inversion in BP.
Towards this end, we turn to the first-principles simulations for
quantitative answers. Our HSE06 calculations show that applying an
external pressure along the $y$ (armchair) or $x$ (zigzag)
direction produces promising effects. Note that, BP is much softer
along the $y$ (armchair) direction because of its anisotropic
atomic structure \cite{2014Peng}. We find that applying a pressure
of 0.75 GPa can shrink the lattice constant along the $y$
direction by 4\% but only by 1\% and 2\% for the $x$ and $z$
directions, respectively. Therefore, we will consider the changes
made by applying pressure along the $y$ direction because it can
change the atomic structure and corresponding electronic structure
more significantly via a relatively small pressure.

First, as shown in Figure 4 (a), an applied critical pressure (0.6
GPa) diminishes the band gap and a 1D Dirac cone is produced along
the $\Gamma-X$ direction. This pressure is well within the
capability of current high-pressure experiments
\cite{2000Katayama, 2003Monaco}.

Second, larger pressures create a band inversion. Excitingly,
similar to the $k \cdot p$ theory predictions, a pressure of 0.9
GPa yields a band crossing at the $\Gamma$ point and a pair of 2D
Dirac cones are formed, as shown in Figure 4 (b). The
corresponding density of states (DOS) is shown in Figure 4 (c). We
can clearly see that a linear DOS in the low-energy regime, which
is similar to graphene. From these first-principles results, we
also find that that the Fermi velocities of compressed BP is
smaller than that of graphene and that these generated Dirac cones
are highly anisotropic in momentum space. Obviously, the Fermi
velocities must be anisotropic as well and can be efficiently
tuned by varying the applied pressure, as concluded in Figure 4
(d).

\begin{figure}
\centering
\includegraphics[scale=0.56]{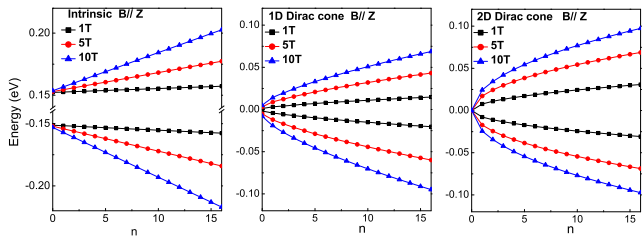}
\caption{(Color online) (a) The energy spectrum of Landau levels
(index n) of intrinsic BP with magnetic field; (b) that under the
critical pressure (0.6 GPa), in which the 1D Dirac cone is formed;
(c) That under the pressure of 0.9 GPa, in which 2D Dirac cones
are formed.} \label{Fig_5}
\end{figure}

We propose that the evolution of the electronic structure from
conventional semiconductors to one possessing 1D Dirac cones and
2D Dirac cones can be observed by measuring the energy spectrum of
Landau levels; this methodology is widely used for identifying
Dirac band structures \cite{2005Zhang}. For example, considering
the X-Y plane, before reaching the critical point for closing the
band gap, the energy spectrum of the Landau levels is that of a
typical semiconductor, i.e., it is linear with the level index, as
shown in Figure 5 (a). Here the spectrum is given by
\begin{equation}
\begin{split}
&E=E_c+(n+1/2)\hbar \omega_c \\
&E=E_v-(n+1/2)\hbar \omega_v,
\end{split}
\end{equation}
where the cyclotron frequencies of electrons and holes are
$\omega_c=e B /\sqrt{m_{cx}m_{cy}}$ and $\omega_v=e B
/\sqrt{m_{vx}m_{vy}}$. $m_{cx}$, $m_{cy}$, $m_{vx}$ and $m_{vy}$
are effective mass of electrons and holes along the $X$ and $Y$
directions, respectively. Importantly, there is no zero-energy
solution for this case.

At the critical pressure, a perfectly 1D linear band dispersion
along the $\Gamma-Y$ direction simplifies Eq. 1 by setting the
parabolic term $\beta$ and $\mu$ to be zero. The effective
Hamiltonian with the magnetic field introduced by the Landau gauge
$\overrightarrow{A}=(-B y, 0, 0)$ is
\begin{equation}
\begin{split}
\frac{1}{2 m_c}(\hbar k_x-e B y)^2 \phi_A -i\hbar v_f \frac{ d}{dy}\phi_B=E \phi_A, \\
-\frac{1}{2 m_v}(\hbar k_x-e B y)^2 \phi_B -i\hbar v_f \frac{
d}{dy}\phi_A=E \phi_B,
\end{split}
\end{equation}
where $ v_f= 3.5 $ x $ 10^{5}m/s$ is the Fermi velocity of the 1D
dirac cone obtained from HSE06 simulations. The corresponding
energy spectrum of Landau Levels in magnetic field is presented in
Figure 5 (b), which is deviated from the linear relation of Figure
5 (a). Finally, for higher pressures that form 2D Dirac cones, the
Landau levels follows the well-known square-root dependence
\cite{2009Neto,2005Zhang}, $E_n \propto \sqrt{n}$ , and exhibits a
unique zero-energy solution, as shown in Figure 5 (c).

In conclusion, we predict that applying a moderate pressure ($>$
0.6 GPa) on bulk BP can dramatically reshape its band structure.
Its band gape will close and band inversion will induce a
topological transition of the Fermi surface via the so-called PSOC
interaction. Under different pressures, the band dispersions of
bulk BP will evolve from that of a semiconductor with a parabolic
band dispersions to a semimetal with 1D or 2D Dirac cones that
exhibit tunable anisotropic Fermi velocities. Compressed BP thus
may be a TI that uniquely integrates both massless Dirac Fermions
and massive classical carriers within one material. Furthermore,
these bulk Dirac-cone states and robust even under broken time
reversal symmetry. Therefore, these unique electronic structures
can be identified by Landau levels and we have provided the
corresponding energy spectra.

We acknowledge fruitful discussions with Li Chen, Zohar Nussinov
and Ryan Soklaski. This work is supported by the National Science
Foundation Grant No. DMR-1207141. The computational resources have
been provided by the Lonestar and Stampede of Teragrid at the
Texas Advanced Computing Center (TACC) and the Edison cluster of
the National Energy Research Scientific Computing Center (NERSC).
The first-principles calculation is performed with the Vienna ab
initio simulation package (VASP) \cite{1996Kresse}.

\end{document}